\documentclass[conference]{IEEEtran}

\usepackage[latin1]{inputenc}
\usepackage{amsmath,bbm,comment,subfigure}
\usepackage[latin1]{inputenc}
\usepackage{amsmath}
\usepackage{amsfonts}
\usepackage{amssymb}
\usepackage{graphicx}
\usepackage{graphics}
\usepackage{lscape}
\usepackage{comment}
\usepackage{mathrsfs}

\usepackage{changes}
\usepackage{lipsum}
\definechangesauthor[name={Authors}, color=blue]{authors}
\setremarkmarkup{(#2)}

\usepackage{acronym}

\usepackage[acronym]{glossaries}
\usepackage{glossary-mcols}
\usepackage{url}
\usepackage{booktabs}
\usepackage{algorithm}
\usepackage[noend]{algpseudocode}
\usepackage{siunitx} 

\def\P{\mathbbm{P}}

\def\E{\mathbbm{E}}
\def\C{\mathbbm{C}}

\newtheorem{lemma}{Lemma}

\newcommand{\QI}{q}

\newtheorem{corollary}{Corollary}

\newtheorem{remark}{Remark}
\newtheorem{hypothesis}{Hypothesis}
\newtheorem{proposition}{Proposition}

\begin{document}

\title{A Processor-Sharing model for the Performance of Virtualized Network Functions}

\author{Fabrice Guillemin, Veronica Quintuna Rodriguez, Alain Simonian \\ 
Orange Labs, France\\email: firstname.lastname@orange.com}

\maketitle

\begin{abstract} 
The parallel execution of requests in a Cloud Computing platform, as for Virtualized  Network Functions, is modeled by an $M^{[X]}/M/1$ Processor-Sharing (PS) system, where each request is seen as a batch of unit jobs. The performance of such paralleled system can then be measured by the quantiles of the batch sojourn time distribution. In this paper, we address the evaluation of this distribution for the $M^{[X]}/M/1$-PS queue with batch arrivals and geometrically distributed batch size. General results on the residual busy period (after a tagged batch arrival time) and the number of unit jobs served during this residual busy period are first derived. This enables us to provide an approximation for the distribution tail of the batch sojourn time whose accuracy is confirmed by simulation. 
\end{abstract} 


\section{Introduction}
Network Function Virtualization (NFV)~\cite{ABD} is deeply modifying the architecture and the operation of telecommunication networks. As a matter of fact, network functions, which were so far hosted on dedicated hardware, are now implemented owing to virtualization technologies on common hardware. A Virtualized Network Function (for short, VNF) is actually most of the time composed of sub-functions which can be executed in parallel or in series. A VNF thus appears as a set of tasks to be executed on a computing system, such as for instance a multi-core platform.

Some sub-functions of a global VNF can be executed in parallel. This is notably the case of the Radio Access Network (RAN)  functions such as the channel coding see~\cite{cloudran}. In this case, it is fundamental to investigate which resource allocation strategy is the most adapted to execute virtualized (sub-)functions on a multi-core platform and, moreover, how cores must be allocated to the tasks that must be executed in parallel. 

From a modelling point of view, a global VNF or a set of tasks, which can be processed in parallel, appears as a batch of unit jobs to be executed on a multi-server system. In this view, batch arrivals correspond to instants when VNF-jobs have to be processed. This leads us to consider multi-server queuing systems with batch arrivals. Several core allocation procedures have been considered by simulation in~\cite{veronica2}, notably the $M^{[X]}/M/C$ and $M^{[X]}/M/1$ Processor Sharing (PS). The analytical study of the $M^{[X]}/M/C$ queue has been performed in~\cite{cloudran}, thus extending earlier results obtained in~\cite{Cromie}. In particular, the asymptotic behavior of the waiting time distribution of an entire batch has been derived.

In this paper, we consider the PS discipline which is a popular method of sharing a common resource between competing tasks. In the context of a multi-core platform, PS consists of sharing the whole computing capacity among all tasks present in the system. The allocation of cores is achieved by the scheduler of the operating system managing the multi-core system. When the PS discipline is performed, all batches are  treated in parallel and  receive equal portion of the computing capacity in a fair basis. It is worth noting that more sophisticated methods could also be envisaged. For instance, the Early Deadline First (EDF) discipline which is often implemented in Linux OS for dealing with real-time applications. However, it is still much more difficult to analyze~\cite{edf}.

In this context, the PS queue with batch arrivals has been already envisaged~\cite{GQS} to calculate the distribution of the sojourn time $W$ of a single job, extending the results obtained by Kleinrock \emph{et al.} for the mean value~\cite{Klein71}. In this paper, we consider the evaluation of the sojourn time $\Omega$ of an entire batch. Although the distribution of $W$ could be given as an explicit integral representation, the exact calculation of the distribution of $\Omega$ proves much more challenging. To overcome this difficulty, we propose an approximation for the distribution tail of $\Omega$ which compares reasonably well to simulation and can be easily handled to quantify the system performance. 
This approximation is derived from (i) general results for the \textit{residual} busy period starting after the arrival time of a tagged batch and (ii) an equi-probability assumption for the departing order of jobs within the residual busy period. 

The analysis performed in this paper allows us to explicitly compute the exponential decay rate of the sojourn time of an entire  batch. This decay rate  globally gives a means of estimating the performance of a resource sharing discipline. In particular, when real time constraints have to be met while executing VNFs (notably in Cloud RAN systems), the decay rate is an indication of the reneging rate of VNFs. In fact, if some VNFs are not executed within prescribed delay bounds, they fail (or renege from a modeling point of view). The results obtained in this paper allow us to compare the PS discipline against the FIFO discipline considered in~\cite{cloudran} for scheduling channel coding tasks in a virtual RAN context.

The paper is organized as follows. Sections \ref{Sec:Intro} and \ref{Sec:RBP} address the exact characterization of the full (resp. residual) busy period of the PS queue with batch arrivals and the number of jobs served during this period. On the basis of the latter results, Section \ref{EDO} then addresses an approximation for the distribution of the batch sojourn time $\Omega$. This approximation is then compared to simulation experiments in Section~\ref{numerical}. Some concluding remarks are finally presented in Section~\ref{conclusion}.

\section{Characteristics of the busy period}
\label{Sec:Intro}

Consider a general $M^{[X]}/G/1$ single server queue with a work-conserving service discipline. This queue is fed by a Poisson process of batches with mean arrival rate $\varrho$; the size (in number of jobs) of any batch is denoted by $B$. The mean service time of a unit job is set equal to 1 and we assume that 
\begin{equation}
\varrho^* \stackrel{def}{=}\varrho \, \E(B) < 1, 
\label{StabCond}
\end{equation}
to ensure the existence of a stationary regime for this queue. Let $T$ be the duration of a busy period; during such a busy period, the number of jobs served is denoted by $M$. After \cite[ Chap.2, Sect.3]{Cohen82}, let 
$\widetilde{F}_m(x) = \mathbb{P}(T \leq x, M = m)$ for $m \geqslant 1$ and  
$x \geqslant 0$, and define the double transform $\nu$ by
$$
\nu(r,s) = \sum_{m \geq 1} r^m \, \int_0^{+\infty} 
e^{-s \, x} \, \mathrm{d}\widetilde{F}_m(x), 
\qquad \vert r \vert < 1, \; s \geqslant 0.
$$
It is known that, given $\vert r \vert < 1$ and $s \geqslant 0$, $\nu(r,s)$ is equal to the smallest root (in modulus) to the equation [ibid., Eq.~(2.15)]
\begin{equation}
\nu = B^*(r \cdot D^*(s + \varrho - \varrho \, \nu))
\label{Equ2}
\end{equation}
where $D^*$ (resp. $B^*$) denotes the Laplace transform of the distribution of the job service time (resp. the generating function of the number of jobs contained in a batch). 

In the rest of this paper, it is assumed that 
\begin{itemize}
\item the identically and independently distributed (i.i.d.) job service times are exponentially distributed with parameter 1 so that $D^*(s) = 1/(1+s)$ for $s \geqslant 0$; 
\item the size (in number of jobs) of a given batch is geometrically distributed with parameter $\QI \in \; ]0,1[$, so that 
\begin{equation}
B^*(z) = \frac{(1-\QI) \, z}{1 - \QI \, z}, \qquad \vert z \vert < 1.
\label{Equ3}
\end{equation}
\end{itemize}
After Eq.~(\ref{Equ3}), in particular, the general  stationarity condition (\ref{StabCond}) now specifies into
\begin{equation}
\varrho < 1 - q.  
\label{StabCondBIS}
\end{equation}
Under these assumptions, we can assert the following.

\begin{lemma}
\textbf{For the $M^{[X]}/M/1$ queue with geometrically distributed batch size, the Laplace transform $T^*$ of the busy period duration $T$ is given by}
\begin{equation}
    \label{Equ4}
T^*(s) = \frac{s+1-q+\varrho -\sqrt{\Delta_q(s)}}{2 \varrho}
\end{equation}
\textbf{for $s \in \C\setminus [\sigma_q^-,\sigma_q^+]$, where}
\begin{equation}
    \label{defDeltaq}
    \Delta_q(s)= ( s+1+\varrho-q)^2 - 4\varrho (1-q) =(s-\sigma_q^+)(s-\sigma_q^-)
\end{equation}
\textbf{and}
\begin{equation}
\label{defsigmapm}
    \sigma_q^\pm = -(\sqrt{1-\QI} \mp \sqrt{\varrho})^2.
\end{equation}
\textbf{The distribution tail of the busy period $T$ decays exponentially fast with rate $\vert \sigma_\QI^+ \vert = (\sqrt{1-\QI} - \sqrt{\varrho})^2$, specifically}
\begin{equation}
\mathbb{P}(T > x) \sim \frac{(1-\QI)^{\frac{1}{4}}}
{2\sqrt{\pi} \, \varrho^{\frac{3}{4}} \, \vert \sigma_\QI^+ \vert} \cdot  
\frac{e^{\sigma_\QI^+ x}}{x^{\frac{3}{2}}}
\label{Equ5}
\end{equation}
\textbf{for large positive $x$.}
\label{L0}
\end{lemma}

\begin{IEEEproof}
Let $T^*(s) = \nu(1,s) = \mathbb{E}(e^{-s \, T})$, $s \geqslant 0$. Applying Eq.~(\ref{Equ2}) for $r = 1$ with $D^*(s) = 1/(1+s)$ and the definition~(\ref{Equ3}) of $B^*$ entails that the Laplace transform $T^*$ verifies
$$
T^*(s) = \frac{1-\QI}{1+s+\varrho - \QI - \varrho \, T^*(s)},
$$
that is,
$\varrho \, T^*(s)^2 - (1+s+\varrho - \QI) \, T^*(s) + 1 - \QI = 0$, 
$s \geqslant 0$, which solves for $T^*(s)$ (equal to the smallest root) into
$$
T^*(s) = \frac{1+s+\varrho-\QI-\sqrt{\Delta_\QI(s)}}{2 \, \varrho}, 
\qquad s \geqslant 0,
$$
with $\Delta_q(s)$ defined by \eqref{defDeltaq}. It is readily verified that $T^*$ then defines an analytic function in the cut plane $\C\setminus [\sigma_-,\sigma_+]$.

Besides, it is known \cite[Sect. 3.46]{Obe73} that the Laplace inverse of transform 
$s \geqslant 0 \mapsto s - \sqrt{s^2 - a^2}$ is 
$t \geqslant 0 \mapsto a \, I_1(at)/t$ for any constant $a > 0$, where $I_1$ 
is the modified Bessel function with order 1; applying the latter inverse with 
$a = 2\sqrt{\varrho(1-\QI)}$, the Laplace inversion of (\ref{Equ4}) entails that the busy period $T$ has the probability density
\begin{equation}
\mathbb{P}(T = t) = 
\sqrt{\frac{1-\QI}{\varrho}} \, \frac{e^{-(1+\varrho - \QI)t}}{t} \,  
I_1(2\sqrt{\varrho(1-\QI)} \cdot t)
\label{GeoT}
\end{equation}
for $t \geqslant 0$. Using the fact that $I_1(X) \sim e^X/\sqrt{2 \pi X}$ for large positive $X$ \cite[Chap.5, Eq.~(5.11.8)]{Leb65}, the tail of this density is therefore asymptotic to
\begin{eqnarray*}
\mathbb{P}(T = t) &\sim& \sqrt{\frac{1-\QI}{\varrho}} \, 
\frac{e^{-(1+\varrho - \QI)t}}{t} \, 
\frac{e^{2 \sqrt{\varrho(1-\QI)}t}}{\sqrt{2\pi \cdot 
2 \sqrt{\varrho(1-\QI)}t}} \\ 
&=& 
\frac{(1-\QI)^{\frac{1}{4}}}{\varrho^{\frac{3}{4}}} \, \frac{e^{\sigma_\QI^+ t}}
{2\sqrt{\pi} \, t^{\frac{3}{2}}}
\end{eqnarray*}
for large positive $t$. The estimate (\ref{Equ5}) of  
$\mathbb{P}(T > x)$ for large positive $x$ follows.
\end{IEEEproof}

Note that the random variable $T$ is the length of the busy period seen by  an external observer. It is not easy to relate the distribution of $T$ to the sojourn time of a batch. In fact, upon arrival, an arbitrary batch sees the system in equilibrium owing to the PASTA property, and not the empty state; we can thus claim that the sojourn time of an arbitrary batch is less than the residual busy period following the batch arrival instant. This sojourn time is further studied in the next section.

\begin{lemma}
\textbf{The generating function $M^*$ of the  number $M$ of customers served in a busy period of the $M^{[X]}/M/1$ queue with geometric batch arrivals is given by}
\begin{equation}
M^*(z)= \frac{1+\varrho -q z-\sqrt{\delta_q(z)}}{2 \varrho}
\label{defM}
\end{equation}
\textbf{for $z \in \C\setminus [\zeta_q^-,\zeta_q^+]$, where}
\begin{equation}
    \label{defdeltaq}
    \delta_q(z) = (1+\varrho-q z)^2-4\varrho  z(1-q)=q^2(z-\zeta_q^-)(z-\zeta_q^+)
\end{equation}
\textbf{and}
\begin{equation}
\label{defzeta}
\zeta_q^\pm = \left ( \frac{\sqrt{\varrho + q} \pm \sqrt{\varrho(1-q)}}{q}\right )^2.
\end{equation}
\textbf{The distribution tail of  $M$ decays exponentially fast with rate $\zeta_q^-$, specifically}
\begin{equation}
\mathbb{P}(M =m) \sim \frac{q\sqrt{(\zeta_q^+-\zeta_q^-)\zeta_q^-}}{4\varrho \sqrt{\pi}} \cdot \frac{1}{m^{\frac{3}{2}}}\left(\frac{1}{\zeta_q^-} \right)^m
\label{asympM}
\end{equation}
\textbf{for large integer $m$.}
\label{LemM}
\end{lemma}

\begin{IEEEproof}
Setting $s=0$ in Eq.~(\ref{Equ2}), we deduce that $M^*(z)$ satisfies
$$
M^*(z) = \frac{(1-q)z}{1+\varrho-\varrho M^*(z)-q},
$$
which quadratic equation has the smallest solution given by expression~\eqref{defM}. 
This equation defines a analytic function in  the cut plane $\C\setminus [\zeta_q^-,\zeta_q^+]$, where $\zeta_q^\pm$ are defined by \eqref{defzeta}. A direct application of Darboux's method \cite[Theorem VI.14]{Flajolet} further yields asymptotics \eqref{asympM}, as claimed.
\end{IEEEproof}

\section{The residual busy period}
\label{Sec:RBP}
As argued in the following, the distribution of the batch sojourn time $\Omega$ for the Processor-Sharing $M^{[X]}/M/1$ queue can be upper bounded to that of the residual busy period after the arrival of a tagged batch. To analyze this residual busy period, we here follow the treatment of \cite[p.249, Section~II.4.4]{Cohen82} for the analysis of the (full) busy period for the $M^{[X]}/G/1$ queue with any work-conserving service discipline. This obtained results will apply, in particular, to the Processor-Sharing discipline subsequently  considered.

\subsection{Joint Laplace transform of $\widetilde{T}$ and $\widetilde{M}$}
\begin{figure}[htb]
\scalebox{0.8}{\includegraphics[width=13cm, trim = 4cm 2cm 0cm 4cm,clip]
{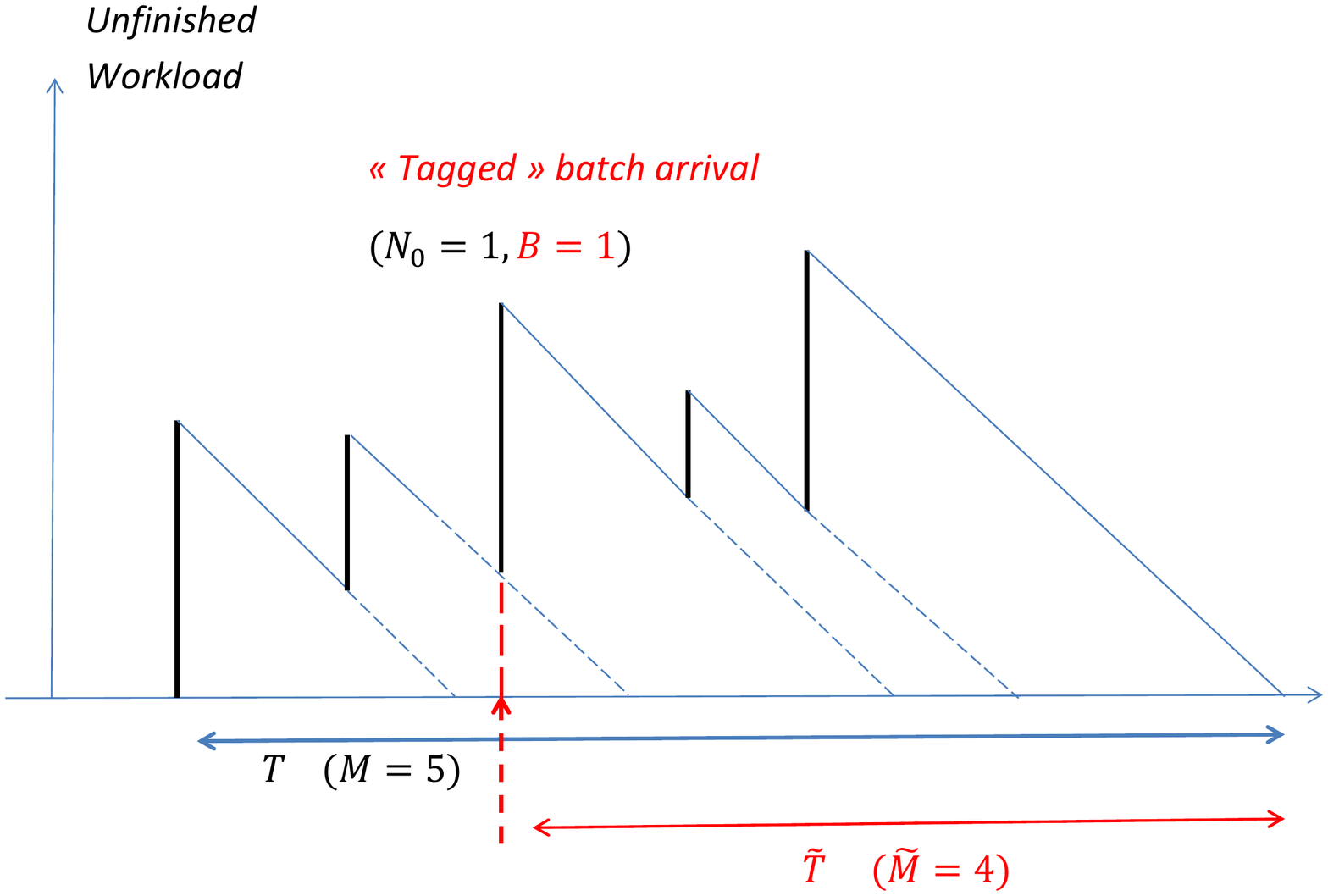}}
\caption{\textit{Busy period $T$ and residual Busy period $\widetilde{T}$.}}
\label{BP}
\end{figure}

Let an $M^{[X]}/M/1$ queue with a work-conserving discipline. Consider a tagged batch with size $B = b$ (in terms of number of jobs) arriving during a busy period; this batch sees a number $N_0 = n \geqslant 0$ of jobs already present in the queue. Let $\widetilde{T}$ (resp. $\widetilde{M}$) further denote the residual duration of the busy period after the arrival time of the test batch (resp. the number of jobs served during this residual duration $\widetilde{T}$). For an illustration, Figure~\ref{BP} displays a sample busy period with duration $T$ and the residual busy period $\widetilde{T}$ associated with the arrival of this tagged batch (recall that for any work-conserving service discipline, the busy period is determined by the smallest interval where the unfinished workload does not reach 0).

Let us introduce the random variable $\widetilde{M}$ equal to the number of jobs served in the residual busy period of length $\widetilde{T}$. We can then state the following result.

\begin{proposition}
\textbf{Given $N_0 = n \geqslant 0$ and $B = b \geqslant 1$, the conditional distribution of the pair $(\widetilde{T}, \widetilde{M})$ is given by}
\begin{equation}
\mathbb{E}_{n,b}(r^{\widetilde{M}} \, e^{-s \, \widetilde{T}}) = 
\left ( \frac{r}{1 + s + \varrho - \varrho \, \nu(r,s)} \right )^{n+b}
\label{Equ8}
\end{equation}
\textbf{for $\vert r \vert < 1$ and $s \geqslant 0$, where $\nu$ is the solution to the functional equation~(\ref{Equ2}).}
\label{P1}
\end{proposition}

\begin{IEEEproof}
Fix $N_0 = n \geqslant 0$, $B = b \geqslant 1$ and define
\begin{itemize}
\item $\tau$ as the cumulative service time of all the $n + b$ jobs present in queue just after the arrival of the tagged batch; 
\item $A_\tau$ as the number of batch arrivals in the time interval with duration $\tau$.
\end{itemize}
\noindent
By definition of $\widetilde{T}$ and $\widetilde{M}$, 
\begin{itemize}
\item[\textbf{(a)}] if $A_\tau = 0$, then
$\widetilde{T} = \tau$ and $\widetilde{M} = n + b$; 
\item[\textbf{(b)}] if $A_\tau \geqslant 1$, with regard  to the duration of the residual busy period, it is indifferent to postpone the service of the remaining $A_\tau - 1$ batches to the end of the busy period generated by the first batch 
$\mathfrak{B}_1$ arrived among this number $A_\tau$. For 
$1 \leqslant k \leqslant A_\tau$, setting then
\begin{itemize}
\item $T_k$ equal to the duration of the busy period generated by the $k$-th batch $\mathfrak{B}_k$ arrived during the interval $\tau$ (with the convention 
$T_0 = 0$),
\item $M_k$ the number of jobs served during the busy period generated by this  
$k$-th batch $\mathfrak{B}_k$ (with the convention $M_0 = 0$),
\end{itemize}
we have the defining equalities 
$\widetilde{T} = \tau + T_1 + ... + T_{A_\tau}$ and 
$\widetilde{M} = n + b + M_1 + ... + M_{A_\tau}$. 
\end{itemize}

By using the above observations for both variables $\widetilde{T}$ and $\widetilde{M}$, the double transform defining the joint distribution of the pair $(\widetilde{T},\widetilde{M})$ then satisfies
\begin{multline*}
\mathbb{E}_{n,b}(r^{\widetilde{M}} \, e^{-s \, \widetilde{T}}) =\\
\mathbb{E}_{n,b}(r^{n + b + M_1 + ... + M_{A_\tau}} \cdot 
e^{-s (\tau + T_1 + ... + T_{A_\tau})})
\end{multline*}
so that
\begin{align}
& \mathbb{E}_{n,b}(r^{\widetilde{M}} \, e^{-s \, \widetilde{T}}) \, = 
\nonumber \\
& r^{n+b} \sum_{k \geq 0} \mathbb{E}_{n,b}(e^{-s \, \tau} 
\mathbf{1}_{A_\tau = k} \; r^{M_1 + ... + M_{A_\tau}} \,  
e^{-s (T_1 + ... + T_{A_\tau})}) \, = 
\nonumber \\
& r^{n+b} \sum_{k \geqslant 0} \int_0^{+\infty} 
\mathrm{d}\mathbb{P}_\tau(t) e^{-s \, t} \, e^{-\varrho \, t} 
\frac{(\varrho t)^k}{k!} \left [ \mathbb{E}(r^M e^{-s T}) \right ]^k
\label{Equ6}
\end{align}
after conditioning with respect to the variable $\tau$,  by noting that the distribution of $A_t$ is Poisson with parameter $\varrho \, t$ and by using the essential fact that all pairs $(T_k,M_k)$, $k \geqslant 1$, are independent and identically distributed. 
Performing the summation with respect to index $k$ in \eqref{Equ6}, we are therefore left with
\begin{align}
& \mathbb{E}_{n,b}(r^{\widetilde{M}} \, e^{-s \, \widetilde{T}}) \; = 
\nonumber \\
& r^{n+b} \int_0^{+\infty} \mathrm{d}\mathbb{P}_\tau(t) e^{-s \, t} \cdot 
e^{-\varrho \, t} \exp \left [ \varrho \, t  \, 
\mathbb{E}(r^M e^{-s T}) \right ] \; = 
\nonumber \\ 
& r^{n+b} \cdot \tau^* (s + \varrho - \varrho \, \mathbb{E}(r^M e^{-s T}))
\label{Equ6bis}
\end{align}
where $\tau^*$ denotes the Laplace transform of variable $\tau$. By the memory-less property of the exponential distribution applied to the remaining service duration of the $n$ jobs present at the arrival instant of the tagged batch, $\tau$ is the sum of $(n + b)$ i.i.d. variables with exponential distribution with parameter $1$; hence
$\tau^*(s) = 1/(1+s)^{n+b}$, $s \geqslant 0$. By equality~(\ref{Equ6bis}) and the latter expression of $\tau^*(s)$, formula (\ref{Equ8}) follows.
\end{IEEEproof}

\subsection{Marginal distributions of $\widetilde{T}$ and $\widetilde{M}$}
By using the joint Laplace transform determined in Proposition \ref{P1}, we now derive the Laplace transform of the duration of the residual busy period and the generating function of the number of jobs served during such a residual busy period.  

First note that the number $N_0$ of jobs present in the queue at the tagged batch arrival instant and the size $B$ of this batch are independent variables. From \cite{GQS}, Eq.~(3.2), we know that the generating function of the number $N_0$ is given by
\begin{equation}
\eta(z) = \mathbb{E}(z^{N_0}) = 
(1-\varrho^*) \, \frac{1-\QI z}{1-(\varrho + \QI)z}, \qquad \vert z \vert < 1,
\label{Eta}
\end{equation}
where $\varrho^* = \varrho/(1-\QI)$.

From the definition~(\ref{Equ3}) of $\mathbb{E}(z^B) = B^*(z)$, the generating function $\varphi$ for the sum $N_0 + B$ is consequently given by 
$\varphi(z)  = \mathbb{E}(z^{N_0+B}) = \eta(z) \, B^*(z)$ which reduces by (\ref{Eta}) to
\begin{equation}
\varphi(z) = \frac{(1-\varrho-\QI)z}{1-(\varrho+\QI)z}, 
\qquad \vert z \vert < 1.
\label{Equ9}
\end{equation}

\begin{proposition}
\textbf{The Laplace transform $\widetilde{T}^*$ of the residual busy period $\widetilde{T}^*$ is given by}
\begin{equation}
    \label{transfoTtilde}
  \widetilde{T}^*(s)  = \frac{(1-q-\varrho)\bigl[-(s+1-\varrho -q)+ \sqrt{\Delta_q(s)} \, \bigr]}{2 \, \varrho \, s}
\end{equation}
\textbf{for $s \in \mathbb{C} \setminus [\sigma_q^-,\sigma_q^-]$, with $\Delta_q(s)$ defined by \eqref{defDeltaq}.}

\textbf{The distribution tail of $\widetilde{T}$ is asymptotic to}
\begin{equation}
\mathbb{P}(\widetilde{T} > x) \sim \frac{(1-q-\varrho)\sqrt{\sigma_q^+-\sigma_q^-}}{4\sqrt{\pi}\varrho (\sigma_q^+)^2} 
\cdot \frac{e^{\sigma_\QI^+ \, x}}{x^\frac{3}{2}}
\label{Equ9bis}
\end{equation}
\textbf{for large $x$, with $\sigma_q^\pm$ defined by \eqref{defsigmapm}.}
\label{C1}
\end{proposition}

\begin{IEEEproof}
By setting $r=1$ in Equation~\eqref{Equ8} and deconditioning on $N_0+B$, we deduce that 
$ \widetilde{T}^*(s)$ is given by
$$
\widetilde{T}^*(s) = 
\varphi \left( \frac{1}{1 + s + \varrho - \varrho \, {T}^*(s)} \right)
$$
with transform $T^*$ given in (\ref{Equ4}) and function $\varphi$ defined in (\ref{Equ9}); simple algebra then provides expression~\eqref{transfoTtilde}, which defines an analytic function in the cut plane  $\C \setminus [\sigma_q^-,\sigma_q^+]$. Besides, (\ref{transfoTtilde}) entails that $\widetilde{T}^*$ has an algebraic singularity at point $\sigma_q^+$ with the expansion
\begin{equation}
\widetilde{T}^*(s) = T_q + 
S_q(s-\sigma_q^+)^{1/2} + o\left ( (s-\sigma_q^+)^{1/2} \right )
\label{DevTtilde}
\end{equation}
when $s\to \sigma_q^+$, where constants $T_q = \widetilde{T}^*(\sigma_q^+)$ and $S_q$ are easily calculated as
\begin{equation}
T_q = 1 + \sqrt{\frac{1-q}{\varrho}}, \quad 
S_q = \frac{(1-q-\varrho)\sqrt{\sigma_q^+-\sigma_q^-}}{2\varrho \sigma_q^+}.
\label{TqSq}
\end{equation}
A  direct application of a classical Tauberian theorem \cite[Theorem 25.2]{Doe58} then yields asymptotics~\eqref{Equ9bis}.
\end{IEEEproof}

Proposition~\ref{C1} has an immediate consequence for the distribution of the batch sojourn time in the $M^{[X]}/M/1$-PS queue. In the sequel, we denote by $\Omega$ the sojourn of an entire batch in this PS queue, that is, the time elapsed between the batch arrival time in queue and the time when all its component jobs have completed their service. 

\begin{corollary}
\textbf{In the \textit{Processor-Sharing} $M^{[X]}/M/1$ queue, the distribution tail of the batch sojourn time $\Omega$ decreases exponentially fast with rate 
$\vert \sigma_\QI^+ \vert$ introduced in  Eq.~(\ref{defsigmapm}).}
\label{C00}
\end{corollary}

\begin{IEEEproof}
As derived in \cite[Cor.~5.2.1]{GQS} for the $M^{[X]}/M/1$-PS queue, the exponential decay rate $\vert \sigma_\QI^+ \vert$ of the distribution of $\widetilde{T}$ (and $T$) equals  that of the distribution of the sojourn time $W$ of a \textit{single job}. 

The inequalities
\begin{equation}
W \leqslant \Omega \leqslant \widetilde{T}, \qquad \mathrm{a.s.},   
\label{Inequ1}
\end{equation} 
then entail that 
$\mathbb{P}(W > x) \leqslant \mathbb{P}(\Omega > x) \leqslant 
\mathbb{P}(\widetilde{T} > x)$ for all $x \geqslant 0$, which enables us to conclude that the distribution tail of the batch sojourn time $\Omega$ also decreases exponentially fast with rate $\vert \sigma_\QI^+ \vert$. 
\end{IEEEproof}

It is worth noting that $\widetilde{T}$ is asymptotically greater than $T$. Indeed, for large $x$, we have
$$
\frac{\P(\widetilde{T}>x)}{\P({T}>x)} \sim \frac{1-\varrho-q}{|\sigma_q^+|}>1.
$$

\begin{proposition}
\textbf{The generating function $\widetilde{M}^*$ of the number $\widetilde{M}$ of jobs served during the residual busy period is given by}
\begin{equation}
\widetilde{M}^*(z) = \frac{(1-q-\varrho) \bigl [ 1 + \varrho - (q + 2\varrho) z - \sqrt{\delta_q(z)} \, \bigr ]}{2\varrho(\varrho+q)(z-1)}
\label{Equ11}
\end{equation}
\textbf{withe $\delta_q(z)$ defined by \eqref{defdeltaq}.}

\textbf{For large $m$, we further have}
\begin{multline}
\P(\widetilde{M}=m) \sim \\   \frac{(1-q-\varrho)q\sqrt{(\zeta_q^+-\zeta_q^-)\zeta_q^-}}{4\sqrt{\pi}\varrho(\varrho+q)(\zeta_q^--1)}  \frac{1}{m^{\frac{3}{2}}}  \left( \frac{1}{\zeta_q^-} \right)^m
\label{Mtildeasym}    
\end{multline}
\textbf{with $\zeta_q^\pm$ defined by \eqref{defzeta}.}
\label{C2}
\end{proposition}

\begin{IEEEproof}
By setting $s = 0$ in Equation~\eqref{Equ8} and deconditioning on $N_0+B$, we deduce that $\widetilde{M}^*(z)$ is given by
$$
\widetilde{M}^*(z) =\varphi\left( \frac{z}{1+\varrho -\varrho M^*(z)} \right)
$$
with generating function $M^*$ given in (\ref{defM}) and function $\varphi$ defined in (\ref{Equ9}); simple algebra then yields expression \eqref{Equ11} for $\widetilde{M}^*(z)$, which defines an analytic function in the cut plane  $\C\setminus[\zeta_q^-,\zeta_q^+]$. When $z$ tends to $\zeta_q^-$, we then derive
\begin{align}
& \widetilde{M}^*(z) = \frac{(1-q-\varrho) (1 + \varrho - (q + 2\varrho) \zeta_q^-)}{2\varrho(\varrho+q)(\zeta_q^--1)} \; - 
\nonumber \\
& \frac{(1-q-\varrho) q\sqrt{(\zeta_q^- - z)(\zeta_q^+-\zeta_q^-)}}
{2\varrho(\varrho+q)(\zeta_q^- - 1)} + o \left ( \sqrt{\zeta_q^- - z} \right )
\nonumber
\end{align}
 and a direct application of Darboux's method provides estimate~\eqref{Mtildeasym}. 
\end{IEEEproof}

Define the sequence $(a_k)_{k \geqslant 0}$ by
$$
a_k = -\frac{1}{(2k-1)2^{2k}}\binom{2k}{k}
$$
so that $\sqrt{1-x}= \sum_{k \geqslant 0} a_k x^k$ for $|x| < 1$  \cite[Eq.(3.6.11)]{Abramowitz}.

\begin{corollary}
\textbf{The distribution of variable $\widetilde{M}$ is given by}
\begin{equation}
\mathbb{P}(\widetilde{M} = m) = 
- \frac{ (1-q-\varrho)(1+\varrho)}{2\varrho(\varrho+q)} \sum_{\ell= m+1}^{+\infty}   b_\ell
\label{Mtilde=m}
\end{equation}
\textbf{for $m \geqslant 1$, where we define}
\begin{equation}
    \label{defbk}
    b_k = \frac{1}{(\zeta_q^-)^k} \sum_{\ell=0}^k a_\ell a_{k-\ell}
    \left( \frac{q\zeta_q^-}{1+\varrho}\right)^{2\ell}, \quad k \geqslant 1.
\end{equation}
\label{C3}
\end{corollary}

\begin{IEEEproof}
Using the fact that $\zeta_q^+\zeta_q^- = (1+\varrho)^2/q^2$, we have
$$
\sqrt{\delta_q(z)} = (1+\varrho)\sqrt{1-\frac{z}{\zeta_q^+}}\sqrt{1-\frac{z}{\zeta_q^-}}
$$
so that $\sqrt{\delta_q(z)} = (1+\varrho) \sum_{k \geqslant 0} b_k z^k$ 
where $b_k$ is defined by \eqref{defbk}. If then follows from (\ref{Equ11}) that 
\begin{multline*}
\widetilde{M}^*(z) = \\  
\frac{(1-q-\varrho)}{2\varrho(\varrho+q)} \times 
\frac{  1+\varrho- (q+2\varrho) z - (1+\varrho)\sum_{k \geqslant 0} b_k z^k  }{z-1}; 
\end{multline*}
by analyticity of $\widetilde{M}^*$, the numerator of the latter fraction vanishes for $z=1$ so that $1-\varrho - q = (1+\varrho)\sum_{k \geqslant 0} b_k$ and thus
$$
\widetilde{M}^*(z) = \frac{(1-q-\varrho)}{2\varrho(\varrho+q)} 
\left[-q  -2\varrho - (1+\varrho)\sum_{k = 1}^{+\infty} b_k \frac{z^k -1} {z-1}\right]. 
$$
By definition of the residual busy period, we have $\widetilde{M} \geqslant 1$ a.s., hence $\widetilde{M}^*(0) = 0$ and the latter power series expansion consequently reduces to
$$
\widetilde{M}^*(z) = \frac{-(1-q-\varrho)(1+\varrho)}{2\varrho(\varrho+q)} 
\sum_{m =1}^{+\infty} z^m \sum_{\ell=m+1}^{+\infty} b_\ell,
$$
whence \eqref{Mtilde=m}. 
\end{IEEEproof}

\section{Estimating the distribution of $\Omega$}
\label{EDO}
While Corollary \ref{C00} has provided us with the exponential decay rate for the distribution of the batch sojourn time $\Omega$ in the Processor-Sharing $M^{[X]}/M/1$ queue, the exact computation of the distribution of $\Omega$ remains, however, extremely challenging. In the present section, we use  results of Sections 
\ref{Sec:Intro} and \ref{Sec:RBP} to propose an approximation for the distribution of sojourn time $\Omega$. 

\begin{figure}[htb]
\begin{center}
\scalebox{0.8}{\includegraphics[width=13cm, trim = 2cm 4cm 1cm 6cm,clip]
{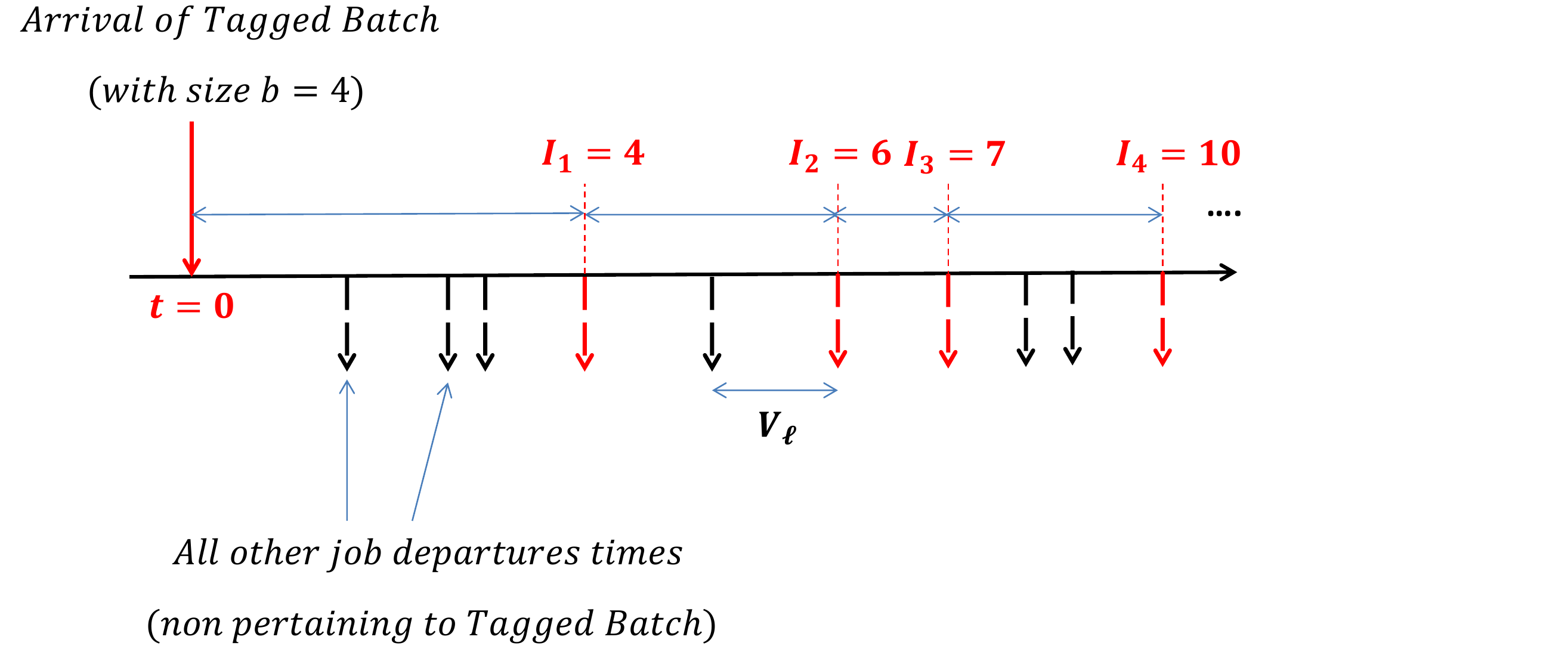}}
\end{center}
\caption{\textit{Consecutive departure times of jobs pertaining to the tagged batch (red arrows) and of jobs non pertaining to this batch (black arrows).}}
\label{TB}
\end{figure}

Let us first introduce a few preliminary definitions. Given the numbers $N_0 = n \geqslant 0$ and $B = b \geqslant 1$, we denote by $I_1 < I_2 < ... < I_b$ the respective departure rank from the queue for each of the $b$ jobs building up the tagged batch; by the above definition of the residual number of jobs served $\widetilde{M}$ after the tagged batch arrival, we certainly have 
\begin{equation}
\forall \; k \in \{1,...,b\}, \quad 1 \leqslant I_k \leqslant \widetilde{M}.
\label{DefI}
\end{equation}
All departure ranks $I_k$, $1 \leqslant k \leqslant b$, being distinct integers by construction, the maximal departure rank $I_b$ also satisfies 
$b \leqslant I_b \leqslant \widetilde{M}$ (see illustration in Fig.\ref{TB}). 

Let $V_\ell$ denote the inter-departure duration between the consecutive completion time of job $\ell-1$ and job $\ell$, $\ell \geqslant 1$, within the residual busy period (jobs $\ell-1$ and $\ell$ may belong to the tagged batch or not). The sojourn time $\Omega$ can then be written as
\begin{equation}
\Omega = V_1 + V_2 + ... + V_{I_b}.
\label{Equ13bis}
\end{equation}
Given the residual number of jobs served $\widetilde{M}$, durations $V_\ell$, 
$1 \leqslant \ell \leqslant I_b$, are dependent random variables: in fact, we have 
$V_1 + ... + V_{I_b} \leqslant \widetilde{T}$ so that, given $\widetilde{M}$ hence 
$\widetilde{T}$, the distribution of the $V_\ell$'s depends on the residual busy period 
and we cannot simply assert that it is exponential with parameter 1.

\subsection{Equi-probability assumption}
\label{EPA}
At this stage, we make the following assumption.
\begin{hypothesis}
\label{hyp1}
\textbf{Given $N_0 = n$, $B = b$ and $\widetilde{M} = m$, all possible departure configurations $(I_1,...,I_b)$ with constraints (\ref{DefI}) for the set of departure ranks of the test tagged are equally probable.}
\end{hypothesis}

Setting then $\xi_k$ as the indicator of the event ``one of the jobs of the tagged batch has departure rank $k$'', $k \in \{1,...,m\}$, Hypothesis 1 entails that
\begin{equation}
\mathbb{P}(\xi_1 = \varepsilon_1, ..., \xi_m = \varepsilon_m) = 
1 \Big / \binom{m}{b}
\label{Equ14}
\end{equation}
for any tuple $(\varepsilon_1,...,\varepsilon_m) \in \{0,1\}^m$ such that 
$\varepsilon_1 + ... + \varepsilon_m = b$; $\binom{m}{b}$ is indeed the number of ways that $b$ balls can be placed into $m \geqslant b$ boxes, each box containing at most one ball. Defining then the variable  
\begin{equation}
\label{defJ}
J = \max \, \left\{j \in \{1,...,m\}, \; \; \;  \sum_{k>j}\varepsilon_k = 0 \right\}
\end{equation}
as the largest box index above which no other boxes contain any ball,  \textbf{Hypothesis~\ref{hyp1}} thus consists in approximating the distribution of largest index $I_b$ by that of $J$. In the following, we study the distribution of random variable $J$.

\begin{lemma}
\textbf{Given $N_0 = n \geqslant 0$, $B = b \geqslant 1$,  
$\widetilde{M} = m \geqslant 1$ and within the equi-probability Hypothesis 1, the conditional distribution of the maximum index $J$ is given by}
\begin{equation}
\mathbb{P}_{n,b,m}(J = j) = \binom{j-1}{b-1} \Big / \binom{m}{b}
\label{Equ15}
\end{equation}
\textbf{for $b \leqslant j \leqslant m$.}
\label{L1}
\end{lemma}

\begin{IEEEproof}
By (\ref{Equ14}) and for $b \leqslant j \leqslant m$, we derive 
$$
\mathbb{P}_{n,b,m}(J \leq j) = \mathbb{P}(\xi_1 + ... + \xi_j = b) = 
\binom{j}{b}  \Big / \binom{m}{b}
$$
so that
$\mathbb{P}_{n,b,m}(J = j) = 
\mathbb{P}_{n,b,m}(J \leq j) - \mathbb{P}_{n,b,m}(J \leq j-1)$ readily reduces to (\ref{Equ15}).
\end{IEEEproof}

The unconditional distribution of the random variable $J$ turns out to be very difficult to compute. We can, nevertheless, estimate its asymptotic behavior at infinity as follows.

\begin{corollary}
\textbf{The unconditional distribution tail of $J$ is given by}
\begin{equation}
    \label{PJ}
    \P(J = j) \sim K_q  \frac{1}{j^{\frac{5}{2}}}\left(\frac{1}{\zeta_q^-}\right)^j  
\end{equation}
\textbf{for large $j$, with constant}
\begin{equation}
    \label{defKq}
K_q = \frac{ \kappa_q\zeta_q^-}{\zeta_q^- - 1} \, 
\frac{(1-\varrho-q)r_q}{(1-qr_q)^2} \, \frac{1-q(\varrho+q)r_q^2}{(1-(\varrho+q)r_q)^2}
\end{equation}
\textbf{where we set}
$$
r_q = \frac{2\zeta_q^-}{1+\varrho +q \zeta_q^-}, \quad 
\kappa_q = \frac{q \sqrt{(\zeta_q^+-\zeta_q^-)\zeta_q^-}}
{ 2\sqrt{\pi}(1+\varrho +q \zeta_q^-)}.
$$
\label{COROL2}
\end{corollary}

The proof of Corollary \ref{COROL2} is detailed in Appendix \ref{App1}.

\subsection{Estimation of the sojourn time of a batch}
\label{AUB}
We now formulate another assumption in order to approximate the distribution tail of sojourn time $\Omega$ of an entire batch. The customers pertaining to a given residual busy period leave the queue after service completion; as mentioned in the introduction of Section \ref{EDO}, we do not actually know the distribution of the inter-departure duration $V_\ell$, $\ell \geqslant 1$; as the queue is work conserving, however, we may reasonably assume that they are independent and identically distributed (recall that this independence assumption can only be an approximation since theses inter-departures are considered conditionally to the fact that they are included in a given residual busy period). This motivates the following assumption.

\begin{hypothesis}
\label{hyp2}
\textbf{The job inter-departure times in a residual busy period are i.i.d.}
\end{hypothesis}

Let then $U$ denote an arbitrary job inter-departure time and $U^*$ its Laplace transform. Given the event  $\widetilde{M}=m$, \textbf{Hypothesis 2} then entails
$U^*(s)^m = E(e^{-s\widetilde{T}}~|~\widetilde{M}=m)$ which, by deconditioning on $\widetilde{M}$ gives $\widetilde{M}^*(U^*(s)) = \widetilde{T}^*(s)$; using the expression (\ref{Equ11}) for $\widetilde{M}^*(z)$, the latter equation readily solves for $U^*(s)$ into
\begin{equation}
\label{deftau}
    U^*(s) = \frac{\bigr [ 1- q -\varrho(q+\varrho)(1- \widetilde{T}^*(s)) \bigr ] \widetilde{T}^*(s)}{R(\widetilde{T}^*(s))}, \; s \geqslant 0,
\end{equation}
where $R(t) = (\varrho t +1 -\varrho - q)((q+\varrho) t + 1 -\varrho - q)$. 

With the above evaluation of the inter-departure time $U$, we now approximate the sojourn time $\Omega$ of a tagged batch of size $b$ as the departure time of the last customer among $b$ customers picked up at random among those customers of the residual busy period. Let $\widetilde{\Omega}$ denote this approximate departure time. 

\begin{hypothesis}
\label{hyp3}
\textbf{Given $J = j \geqslant b$ and following (\ref{Equ13bis}), the distribution of the sojourn time $\Omega$ is approximated by the sum 
$\widetilde{\Omega} = U_1 + U_2 + ... + U_j$ where the $U_\ell$'s are i.i.d. random variables with the distribution of $U$ defined by Eq.~\eqref{deftau}.}
\end{hypothesis}

Invoking \textbf{Hypothesis 2} and  \textbf{Hypothesis 3} now enable us to obtain the following evaluation for the distribution tail of $\Omega$.

\begin{proposition}
\textbf{The distribution tail of the sojourn time $\Omega$ of of an entire batch can be approximated by}
\begin{equation}
   \mathbb{P}(\widetilde{\Omega} > x) \sim \frac{H_qL_q}{2 \sigma_q^+ \sqrt{\pi}} 
   \cdot 
   \frac{e^{\sigma_q^+ x}}{x^\frac{3}{2}}
    \label{ApproxOmega}
\end{equation}
\textbf{for large $x$, with multiplying factor}
$$
H_q = \frac{\mathrm{d}J^*}{\mathrm{d}z}(U^*(\sigma_q^+))
$$
\textbf{where $J^*$ denotes the generating function of variable $J$ and with argument}
\begin{equation}
U^*(\sigma_q^+) = \frac{1+ \varrho - \sqrt{\varrho(1-q)}}{q+\sqrt{\varrho(1-q)}},
\label{U*Sigmaq}
\end{equation}
\textbf{along with}
\begin{equation}
L_q =\frac{\sigma_q^+\sqrt{\sigma_q^+-\sigma_q^-}}{2(q+\sqrt{\varrho(1-q)})^2}. 
\label{defLq}
\end{equation}
\label{PropOmegaTilde}
\end{proposition}

\begin{IEEEproof}
Following \textbf{Hypothesis 3}, the Laplace transform of $\widetilde{\Omega}$ is given by
\begin{equation}
\E(e^{-s \, \widetilde{\Omega}}) = J^*(U^*(s)), \qquad s \geqslant 0.
\label{DefOmegaTild}
\end{equation}
We claim that the smallest singularity of transform  (\ref{DefOmegaTild}) in the complex plane is algebraic and located at $s = \sigma_q^+$. In fact, we make the following points:

$\bullet$ After (\ref{TqSq}), the value $\widetilde{T}^*(\sigma_q^+) = T_q$ is finite and positive. Besides, the function $s \mapsto \widetilde{T}^*(s)$ decreases on the real interval $[\sigma_q^+,+\infty[$ from $T_q > 0$ to 0. In fact, we calculate
\begin{multline*}
\frac{\mathrm{d}\widetilde{T}}
{\mathrm{d}s}(s) =\frac{1-q-\varrho}{2\varrho s^2\sqrt{\Delta_q(s)}} \, \times \\ \left[-(1-q-\varrho)^2 - s(1-q+\varrho) + (1-q-\varrho)\sqrt{\Delta_q(s)} \right].
\end{multline*}
For $s>\sigma_q^+$, we have (*)  $(1-q-\varrho)^2+s(1-q+\varrho)>0$ (if this quantity were  negative, we would have  $s<-(1-q-\varrho)^2/(1-q+\varrho)$; but the inequality $(\sqrt{1-q}+\sqrt{\varrho})^2>1-q+\varrho$ implies in turn  $-(1-q-\varrho)^2/(1-q+\varrho)<\sigma_q^+$ and then $s<\sigma_q^+$, a contradiction). It follows that for $s>\sigma_q^+$, inequality (*) and the identity
\begin{multline*}
(1-q-\varrho)^2{\Delta_q(s)} - \left((1-q-\varrho)^2+s(1-q+\varrho)\right)^2 \\  
= -4(1-q)\varrho s^2
\end{multline*}
imply that $\mathrm{d}\widetilde{T}(s)/\mathrm{d}s \leqslant  0$ for $s>\sigma_q^+$ and the function $\widetilde{T}$ is monotonic decreasing on $[\sigma_q^+,+\infty[$, as claimed.

From definition (\ref{deftau}), polynomial $R(t)$ has negative roots which cannot therefore be attained by $T^*(s) \geqslant 0$, $s \geqslant \sigma_q^+$.  We conclude that $R(T^*(s))$ cannot vanish on this interval. As being well-defined on interval $[\sigma_q^+,+\infty[$, the Laplace transform $U^*$ introduced in Eq.~(\ref{deftau}) is thus well-defined over the whole half-plane $\{s \in \mathbb{C}, \; \Re(s) \geqslant \sigma_q^+\}$.


$\bullet$ By Proposition \ref{COROL2}, the generating series $J^*(z)$ is convergent for $\vert z \vert < \zeta_q^-$. We further verify that the value $U^*(\sigma_q^+)$ of the argument of $J^*$ in (\ref{DefOmegaTild}) for $s = \sigma_q^+$ is less than this convergence radius $\zeta_q^-$. In fact, expression (\ref{deftau}) and simple algebra easily provide formula (\ref{U*Sigmaq}) given in the Proposition for $U^*(\sigma_q^+)$. It is then first easily checked that $U^*(\sigma_q^+)>1$; in addition, the difference 
\begin{multline*}
\zeta_q^- -U^*(\sigma_q^+) = \\  \frac{\sqrt{\varrho(1-q)}}{q^2(q+\sqrt{(1-q)\varrho})} 
\left[ q+ \sqrt{(1-q)\varrho}-\sqrt{\varrho+q} \right]^2
\end{multline*}
is non negative and vanishes for $\varrho = 1 - q$ only, which is excluded by the stability condition (\ref{StabCondBIS}); this consequently shows that $1<U^*(\sigma_q^+)<\zeta_q^-$, as claimed.

Setting $U^*(s) = \mathcal{U}(T^*(s))$ for short and using expansion (\ref{DevTtilde}) for $\widetilde{T}^*(s)$, we then have
\begin{equation}
U^*(s) = \mathcal{U}(T_q) + 
L_q(s-\sigma_q^+)^{1/2} + o \left ( (s-\sigma_q^+)^{1/2} \right )
\label{DevU}
\end{equation}
in the neighborhood of the singularity $s = \sigma_q^+$, where we set 
$L_q = \mathcal{U}'(T_q) S_q$ with constants $T_q$ and $S_q$ given in (\ref{TqSq}). We calculate $\mathcal{U}'(t) = (1-\varrho-q)^2(1-q-\varrho(q+\varrho)(1-t)^2)/R(t)^2$ 
so that
$$
\mathcal{U}'(T_q) = \frac{\varrho(\sigma_q^+)^2}{(1-q-\varrho)(q+\sqrt{\varrho(1-q)})^2}
$$
hence the explicit expression (\ref{defLq}) given in the Proposition for 
$L_q = \mathcal{U}'(T_q) S_q$. By expansion (\ref{DevU}), transform (\ref{DefOmegaTild}) consequently expands at first order in $(s-\sigma_q^+)^{1/2}$ as
\begin{align}
\E(e^{-s \, \widetilde{\Omega}}) = & \, 
J^* \left( \mathcal{U}(T_q) + L_q(s-\sigma_q^+)^{1/2} + ... \right )
\nonumber \\
= & \, J^*(U^*(\sigma_q^+)) + H_q L_q (s-\sigma_q^+)^{1/2} + ...
\nonumber
\end{align}
where $H_q = \mathrm{d}_z J^*(U^*(\sigma_q^+))$ denotes the first derivative of $J^*$ at point $U^*(\sigma_q^+)$. Applying the Tauberian theorem \cite[Theorem 25.2]{Doe58} then provides estimate (\ref{ApproxOmega}), as claimed.
\end{IEEEproof}

\section{Numerical results}
\label{numerical}

To validate the accuracy of the propositions asserted in the previous sections, we simulate a Processor-Sharing system where jobs have exponentially distributed service times with unit mean and arrive in batches with geometrically distributed size with parameter $q$,  according to a Poisson process with rate $\varrho$ such that $\varrho^* = \frac{\varrho}{1-q}
<1$. We simulate batches arriving to the system in equilibrium. We have simulated more than $10^7$ batches to compute distributions of random variables $\Omega$ and $I_b$ as well as the associated random variable $J$.

\begin{figure}[htb]
\centering
\scalebox{0.66}{\includegraphics[width=11cm, trim = 3cm 9cm 3cm 9cm,clip]
{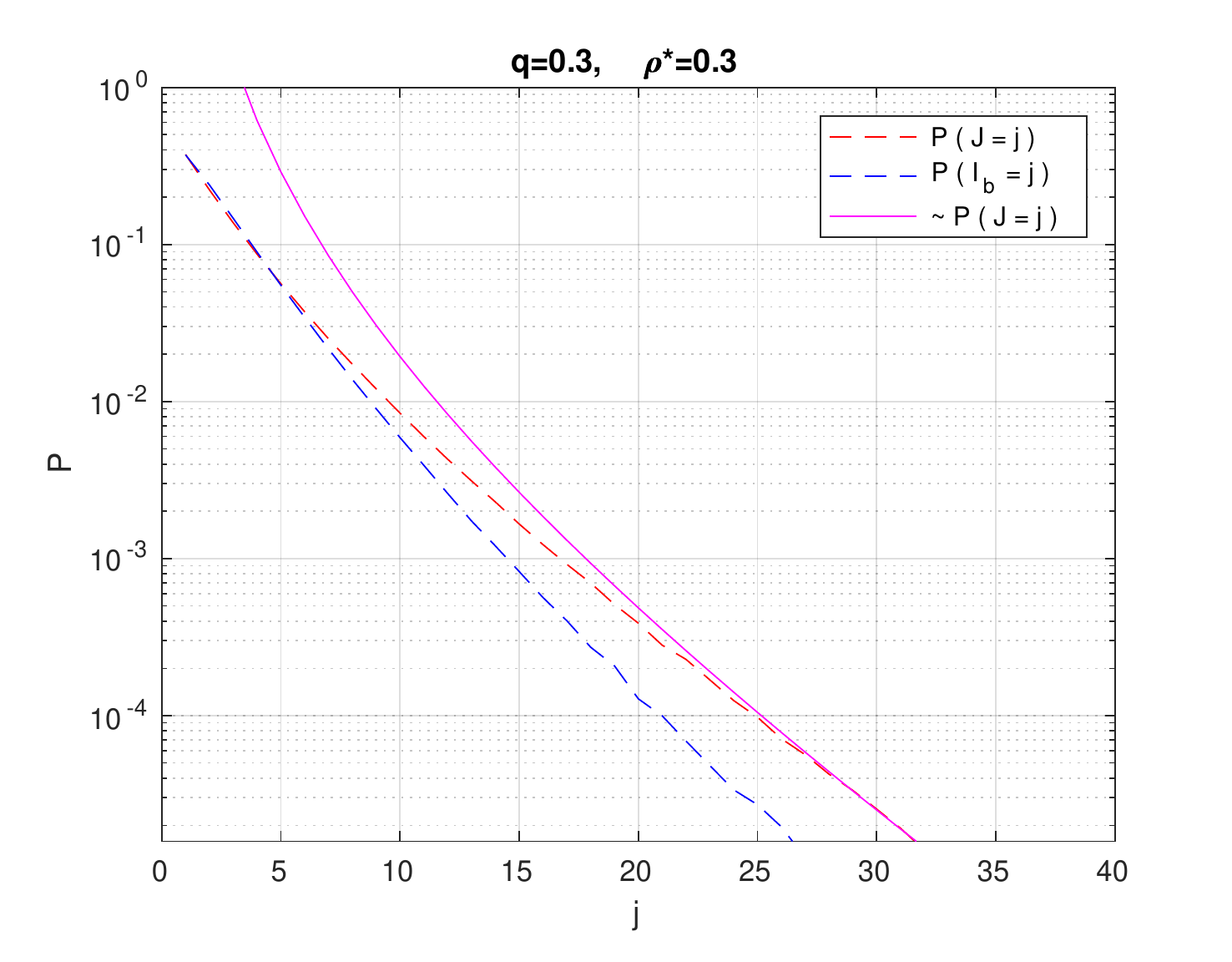}}
\caption{\textit{P(J=j) for $q=0.3$, $\varrho^*=0.3$}}
\label{figJ33}
\end{figure}

\begin{figure}[htb]
 \centering
\scalebox{0.66}{\includegraphics[width=11cm, trim = 3cm 9cm 3cm 9cm,clip]
{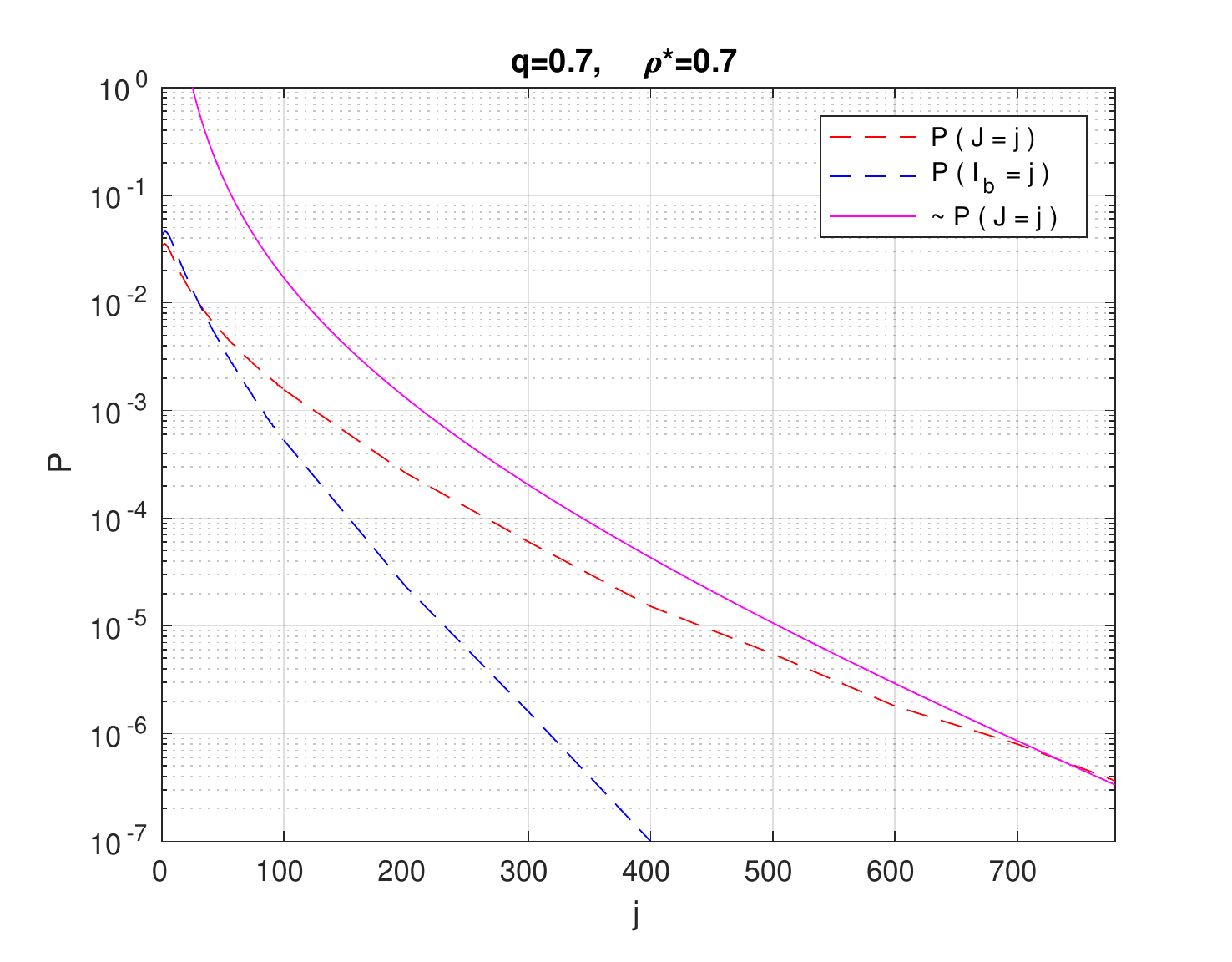}}
\caption{\textit{P(J=j) for $q=0.7$, $\varrho^*=0.7$}}
\label{figJ77}
\end{figure}

In a first step, we examine the equi-probability Hypothesis~\ref{hyp1}. We compare the index of the last job of the tagged batch leaving the system (denoted by $I_b$) to the index $J$ computed by randomly picking up a number of jobs equal to the size of the tagged batch. In Figures~\ref{figJ33} and \ref{figJ77}, we plot the probability density distribution of these two random variables as well as the approximation given by Equation~\eqref{PJ}.

 In Figure~\ref{figJ33}, the load of the system and the mean batch size are rather small and the proposed approximation is quite accurate. In  Figure~\ref{figJ77}, we increase the load and the batch size;  the proposed approximation is still relevant for small indexes but becomes loose for larger  ones. Nevertheless, we empirically observe that the proposed approximation yields an upper bound for the index of the last job of the tagged batch leaving the system.
 
 We now consider the sojourn time $\Omega$. Because of Hypotheses~\ref{hyp2} and \ref{hyp3}, the random variable  $\widetilde{\Omega}$ cannot be easily estimated because the probability distribution of  inter-departure times of jobs within a busy period is not known. Instead, we introduce another random variable  $\hat{\Omega}$ equal to the departure time of the last batch, when picking up at random a number of jobs equal to the batch size and when setting the time origin equal to the tagged job arrival time.
 
\begin{figure}[htb]
\centering
\scalebox{0.66}{\includegraphics[width=11cm, trim = 3cm 9cm 3cm 9cm,clip]
{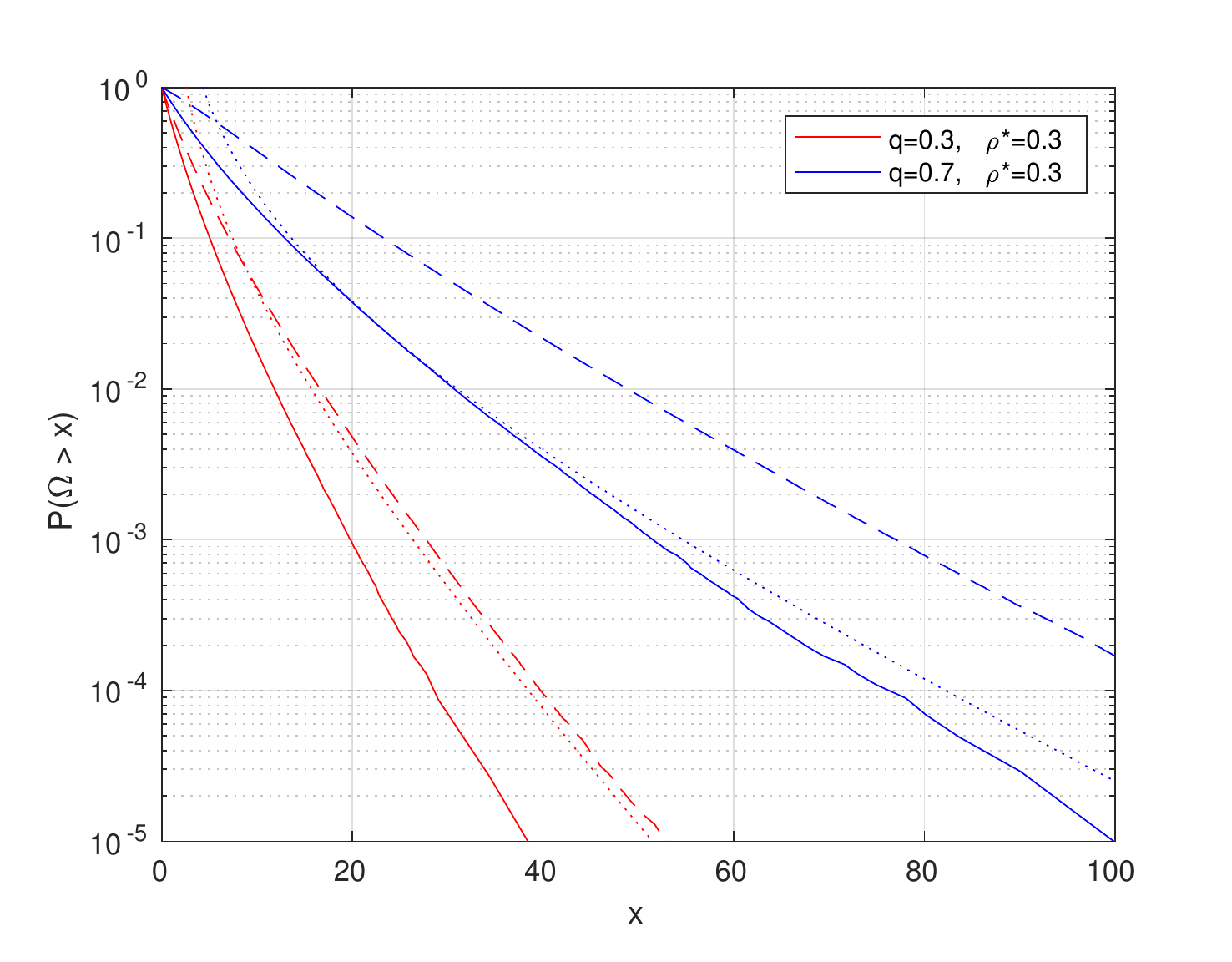}}
\caption{\textit{$\Omega$ solid line, $\hat{\Omega}$ dashed line, $\widetilde{\Omega}$ dotted line}}
\label{figomega1}
\end{figure}

 In Figure~\ref{figomega1}, we  plot the complementary cumulative distribution function of random variables $\Omega$ and  $\hat{\Omega}$ for a light load and for both small and  moderate mean batch size. We observe that the approximation is reasonably accurate. We have also represented approximation~\eqref{ApproxOmega} for $\widetilde{\Omega}$. For computing the multiplying  factor
 $$
 H_q =\sum_{j=1}^{+\infty} \, j \P(J=j) \left(\frac{U^*(\sigma_q^+)}{\zeta_q^-}  \right)^{j-1}
 $$
 introduced in (\ref{ApproxOmega}), we use the values of $ \P(J=j)$, $j \geqslant 1$,  obtained by simulation. It turns out that this approximation is much better than $\hat{\Omega}$ for large values of the mean batch size. The random variable $\hat{\Omega}$ is easy to simulate but difficult to study analytically while it is exactly the contrary for $\widetilde{\Omega}$.  
 
 As previously observed for the evaluation of variable $J$, the approximation is reasonably accurate for small values of the system load but becomes less accurate for larger values. As observed earlier, approximation~\eqref{ApproxOmega} yields better results.
 
 \begin{figure}[htb]
 \centering
\scalebox{0.66}{\includegraphics[width=11cm, trim = 3cm 9cm 3cm 9cm,clip]{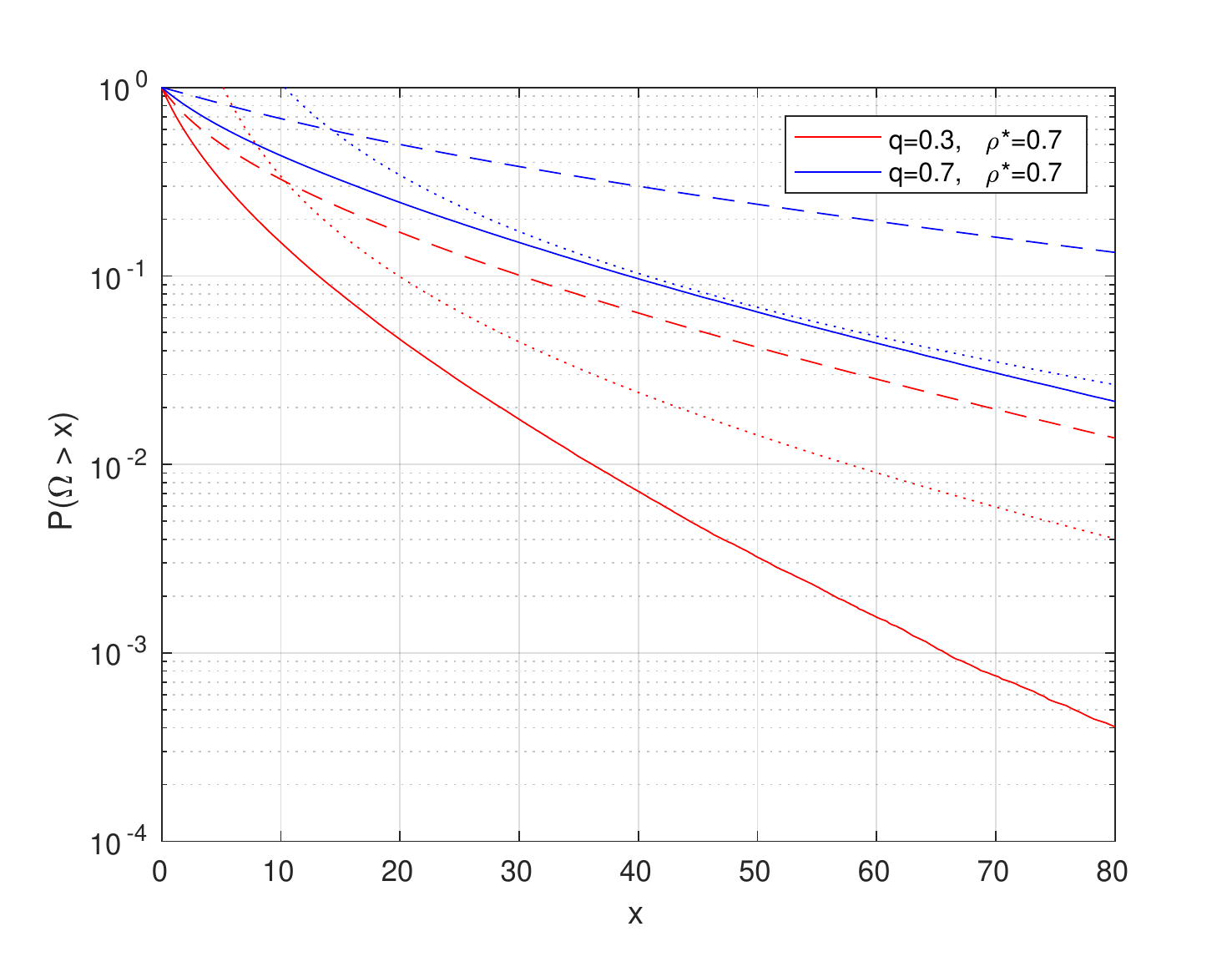}}
\caption{\textit{$\Omega$ solid line, $\hat{\Omega}$ dashed line, $\widetilde{\Omega}$ dotted line}}
\label{figomega2}
\end{figure}

\section{Conclusion}
\label{conclusion}

In this paper, we have considered the sojourn time of an entire batch in the $M^{[X]}/M/1$-PS system. Since this quantity is difficult to study analytically, we have introduced two approximations, one for the index of the last job of a tagged batch leaving the system (the index is obtained by labeling jobs according to their departure instants after the batch arrival) and another  for the sojourn time of the entire batch. Simulations show that the proposed approximations give reasonable results.

From a practical point of view, we conclude from the computations carried out in this paper that the decay rate of the sojourn time of batch in the  $M^{[X]}/M/1$-PS system is $|\sigma_q^+|$ defined by Equation~\eqref{defsigmapm}. By using results from~\cite{cloudran}, we can easily see that this decay rate is less than the one associated with the $M^{[X]}/M/C$ queue. Hence, if we introduce deadlines in the execution of VNFs, the rate of overrun will be higher in the   $M^{[X]}/M/1$-PS than in the $M^{[X]}/M/C$ system. This confirms the earlier results obtained in~\cite{veronica2} by simulation.
\bibliographystyle{IEEEtran}
\bibliography{biblio}
\appendix
\section{Appendix}
\subsection{Proof of Corollary \ref{COROL2}}
\label{App1}
After the identity \cite[Equ.~(6.2.2)]{Abramowitz}
$$
 \int_0^1 t^{\alpha-1}(1-t)^{\beta-1} \mathrm{d}t =  \frac{\Gamma(\alpha)\Gamma(\beta)}{\Gamma(\alpha+\beta)},
$$ 
write
$$
1 \Big / \binom{m}{b} = b \int_0^1 t^{b-1}(1-t)^{m-b} \, \mathrm{d}t.
$$
Consequently, expression (\ref{Equ15}) equivalently reads
$$
\P_{n,b,m}(J =  j) = \frac{(j-1)!}{(b-1)!(j-b)!} 
\int_0^1 \frac{b \, t^{b-1}}{(1-t)^b}(1-t)^m \, \mathrm{d}t
$$
and deconditioning with respect to variable $\widetilde{M}$ gives
\begin{align}
& \, \P_{n,b}(J =  j) = \frac{(j-1)!}{(b-1)!(j-b)!} \; \times 
\nonumber \\ 
& \, \int_0^1 \frac{b \, t^{b-1}}{(1-t)^b}\sum_{m=j}^{+\infty} (1-t)^m  \P_{n,b}(\widetilde{M}=m) \, \mathrm{d}t.
\label{PJ0}
\end{align}
We now evaluate $\P_{b,n}(\widetilde{M}=m)$ for large $m \geqslant j$. Using the expression (\ref{defM}) of $M^*(z)$ with smallest singularity located at $z = \zeta_q^-$, we first have
$$
M^*(z) = \frac{1+\varrho -q\zeta_q^--q\sqrt{(\zeta_q^+-\zeta_q^-)(\zeta_q^--z)}}{2\varrho} + ...
$$
where dots denote $o \, ( \sqrt{\zeta_q^- - z} \, )$ terms when $z \rightarrow \zeta_q^-$; applying relation \eqref{Equ8} to $s = 0$ and with $\nu(z,0) = M^*(z)$, we deduce that
\begin{align}
\E_{n,b}(z^{\widetilde{M}}) = & \, r_q^{n+b} \, \times
\nonumber \\
& \left [ 1 - (n+b) \frac{q\sqrt{(\zeta_q^+-\zeta_q^-)\zeta_q^-}}{(1+\varrho +q \zeta_q^-)} \sqrt{1-\frac{z}{\zeta_q^-}} + ... \right ]
\nonumber
\end{align}
when $z \rightarrow \zeta_q^-$, where we set 
$r_q = 2\zeta_q^-/(1 + \varrho + q \, \zeta_q^-)$ for short. A direct application of Darboux's method \cite[Theorem VI.14]{Flajolet} then yields the asymptotics 
$$
\P_{n,b}(\widetilde{M} = m) \sim \kappa_q (n+b) r_q^{n+b} 
\frac{1}{m^{\frac{3}{2}}} \left(\frac{1}{\zeta_q^-}\right)^m
$$
for large $m$, with constant $\kappa_q$ set as in (\ref{defKq}). Using the latter estimate of $\mathbb{P}_{n,b}(\widetilde{M} = m)$, we consequently deduce that
\begin{multline*}
\sum_{m=j}^{+\infty} (1-t)^m  \P_{b,n}(\widetilde{M}=m) \sim \\
\frac{\kappa_q \zeta_q^-(n+b)r_q^{n+b}}{\zeta_q^- - 1 + t}
\frac{1}{j^{\frac{3}{2}}}\left(\frac{1-t}{\zeta_q^-}\right)^j
\end{multline*}
for large $j$ so that expression (\ref{PJ0}) yields in turn 
\begin{align}
& \, \P_{n,b}(J =  j) \sim (n+b) r_q^{n+b} \frac{1}{j^{\frac{3}{2}}}\left(\frac{1}{\zeta_q^-}\right)^j \; \times 
\nonumber \\ 
& \, \frac{b}{(b-1)!} \int_0^1 (j t)^{b-1}(1-t)^{j-b} 
\frac{\kappa_q \zeta_q^-}{\zeta_q^- - 1 + t} \, \mathrm{d}t
\label{PJ1}
\end{align}
where we have used the fact that $(j-1)!/(j-b)!\sim j^{b-1}$ for large $j$ and fixed $b$. To finally evaluate the integral appearing in (\ref{PJ1}) for large $j$, the variable change $u = j t$ simply provides
\begin{align}
& \frac{b}{(b-1)!} \int_0^1 (j t)^{b-1}(1-t)^{j-b} 
\frac{\kappa_q \zeta_q^-}{\zeta_q^- - 1 + t} \, \mathrm{d}t \, = 
\nonumber \\
& \frac{b}{j (b-1)!} \, \int_0^j u^{b-1} \left( 1-\frac{u}{j} \right)^{j-b}
\frac{\kappa_q \zeta_q^-}{\zeta_q^--1+u/j} \, \mathrm{d}u \, \sim 
\nonumber \\
& \frac{b \kappa_q\zeta_q^-}{j(\zeta_q^- - 1)} \frac{\Gamma(b)}{(b-1)!} = 
\frac{b \kappa_q\zeta_q^-}{j(\zeta_q^- - 1)}
\nonumber
\end{align}
by definition of the Euler $\Gamma$ function; using the latter and estimate (\ref{PJ1}), 
we deduce
\begin{equation}
\P_{n,b}(J =  j) \sim \frac{ \kappa_q\zeta_q^-}{\zeta_q^--1}    \frac{1}{j^{\frac{5}{2}}}\left(\frac{1}{\zeta_q^-}\right)^j \cdot  
b(n+b) r_q^{n+b}
\label{PJ2}
\end{equation}
for large $j$. We finally note that 
\begin{equation}
r_q = \frac{2\zeta_q^-}{1+\varrho +q \zeta_q^-} < \frac{1}{\varrho+q};
\label{Ineqtech}
\end{equation}
in fact, calculating
$$
    \label{ineqtech}
    \delta_q\left(\frac{1+\varrho}{2\varrho+q}\right) = -\frac{4 \varrho(1+\varrho)(1-q-\varrho)(q+\varrho)}{(q+2\varrho)^2}  < 0
$$
together with condition (\ref{StabCondBIS}) show that
$$
\frac{1+\varrho}{2\varrho+q} > \zeta_q^-
$$
hence inequality~\eqref{Ineqtech};  this consequently ensures that 
$\E(r_q^{N_0 + B}) < +\infty$ after (\ref{Equ9}). Deconditioning each side of (\ref{PJ2}) on variables $N_0$ and $B$ then provides asymptotics \eqref{PJ}, with associated constant
\begin{equation}
K_q = \frac{ \kappa_q\zeta_q^-}{\zeta_q^--1} \cdot 
\mathbb{E} \left [ B(N_0 + B) r_q^{N_0 + B} \right ].
\label{defKq0}
\end{equation}
Using the respective definitions (\ref{Equ3}) and (\ref{Eta}) of generating function $B^*$ and $\eta^*$, it is easily verified that the expectation in (\ref{defKq0}) equals 
\begin{align}
& \mathbb{E} \left [ B(N_0 + B) r_q^{N_0 + B} \right ] \; = 
\nonumber \\
& r_q^2 \frac{\mathrm{d}B^*}{\mathrm{d}z}(r_q) \, 
\frac{\mathrm{d}\eta^*}{\mathrm{d}z}(r_q) + 
r_q \left ( \frac{\mathrm{d}B^*}{\mathrm{d}z}(r_q) + r_q \frac{\mathrm{d}^2B^*}{\mathrm{d}z^2}(r_q)\right ) \eta(r_q);
\nonumber
\end{align}
the latter together with (\ref{defKq0}) yield the final expression (\ref{defKq}) of constant $K_q$ after simple algebra. $\blacksquare$

\end{document}